\documentclass[a4paper, 12pt]{article}
\usepackage{amsfonts, amsmath, amssymb, epsfig, subfigure, graphicx}

\usepackage[toc,page]{appendix}
\usepackage{authblk}

\usepackage{tikz}
\usepackage{tabularx,ragged2e,booktabs,caption}
\usetikzlibrary{calc,patterns}
\usepackage{xcolor}
\newcommand{\commentout}[1]{}
 \footnotesep=\baselineskip
\def\wl{\par \vspace{\baselineskip}}

\usepackage{fancyhdr}
\begin{document}
\author{Laura Biziou-van-Pol$^1$, Jana Haenen$^1$, Arianna Novaro$^1$, Andr\'es Occhipinti Liberman$^1$, Valerio Capraro$^2$}
\affil{$^1$Institute for Logic, Language, and Computation, Universiteit van Amsterdam, 1090 GE, Amsterdam, The Netherlands. $^2$Center for Mathematics and Computer Science, 1098 XG, Amsterdam, The Netherlands. \wl Contact Author: V.Capraro@cwi.nl}
\title{Does telling white lies signal pro-social preferences?}
\maketitle

\begin{center}
\emph{Forthcoming in Judgment and Decision Making}
\end{center}

\pagebreak

\begin{abstract}
The opportunity to tell a white lie (i.e., a lie that benefits another person) generates a moral conflict between two opposite moral dictates, one pushing towards telling always the truth and the other pushing towards helping others. Here we study how people resolve this moral conflict. What does telling a white lie signal about a person's pro-social tendencies? To answer this question, we conducted a two-stage 2x2 experiment. In the first stage, we used a Deception Game to measure aversion to telling a Pareto white lie (i.e., a lie that helps both the liar and the listener), and aversion to telling an altruistic white lie (i.e., a lie that helps the listener at the expense of the liar). In the second stage we measured altruistic tendencies using a Dictator Game and cooperative tendencies using a Prisoner's dilemma.  We found three major results: (i) both altruism and cooperation are positively correlated with aversion to telling a Pareto white lie; (ii) both altruism and cooperation are negatively correlated with aversion to telling an altruistic white lie; (iii) men are more likely than women to tell an altruistic white lie, but not to tell a Pareto white lie. Our results shed light on the moral conflit between pro-sociality and truth-telling. In particular, the first finding suggests that a significant proportion of people have non-distributional notions of what the right thing to do is: irrespective of their economic consequences, they tell the truth, they cooperate, they share their money.
\end{abstract}


\pagebreak
\section{Introduction} 

\par Moral decision-making in communication often concerns the choice whether to tell the truth or to deceive. Whilst it is generally agreed that it is bad to tell lies that increase your benefit at the expense of that of another person (black lies), moral philosophers have long argued about if and when telling a lie that increases the benefit of another person (white lie) is morally acceptable. 
We find Socrates pointing out to one of his interlocutors in Plato's Republic that, ``when any of our so-called friends are attempting, through madness or ignorance, to do something bad, isn't it a useful drug for preventing them?'' - suggesting that, given the circumstances, deception might be the `good' thing to do (Plato, 1997). At the other end of the spectrum we find Immanuel Kant, for whom good intentions or consequences cannot justify an act of lying. For Kant, telling even a white lie is ``by its mere form a crime of a human being against his own person and a worthlessness that must make him contemptible in his own eyes.'' (Kant, 1996). 

\par This raises the question whether prosocial agents would tell such `useful' lies, or condemn them, as Kant did. Prosocial behaviour, that is, behaviour intended to benefit other people or society as a whole, is widely considered as the right course of action in situations in which there is a conflict between one's own benefit and that of others. The Golden Rule, which encapsulates the essence of prosociality, is indeed ``found in some form in almost every ethical tradition'' (Blackburn, 2003). Thus, a prosocial person, when facing the decision of whether to tell a white lie or not may experience a conflict between two diverging moral dictates, one pushing towards lying for the benefit of others and the other pushing towards telling the truth regardless of circumstance.

\par Since most human interaction revolves around communication and involves some degree of prosociality, understanding how this conflict is resolved is not only interesting from the theoretical point of view of moral philosophy, but also from a more practical point of view. For instance, taking verbatim an example from Erat and Gneezy (2012): ``should a supervisor give truthful  feedback  to a poorly performing employee, even when such truthful feedback has the potential to reduce the employee's confidence and future performance?'' What does telling a white lie signal about the supervisor's prosocial tendencies? 

\par The focus of the present paper is on this type of questions and, more generally, the moral conflict between lying aversion and prosocial behaviour. 

\par To measure prosocial tendencies and aversion to telling white lies we build on previous studies in behavioural economics, which place economic games into experimental settings.  Specifically, the Dictator Game (DG), due to its setup, has proven useful in measuring altruistic proclivities in recruited subjects. In a standard DG, one player (the dictator) is given an initial endowment and is asked to decide how much of it, if any, to transfer to a passive player (the recipient), who is given nothing. The anonymity and confidentiality of decisions are ensured to rule out incentives (such as reputation) to share their endowment with the recipient. Although the theory of homo \oe conomicus predicts that dictators keep the whole endowment for themselves, research has shown that a significant proportion of dictators allocate a non-trivial share to recipients (Camerer 2003; Engel 2011; Forsythe, Horowitz, Savin \& Sefton, 1994; Kahneman, Knetsch \& Thaler, 1986). 

\par

Akin to the manner in which the DG is used in research as the paradigmatic game with which to
investigate altruism, extensive use has been made of the Prisoner's
Dilemma (PD) in experimental settings in order to investigate
cooperative behaviour in agents. In the standard one-shot two-player
PD, both players can either cooperate or defect. If a player
cooperates, he pays $c$ and bestows $b>c$ on the other player while,
if he defects, he pays and gives $0$. Clearly homo \oe conomicus
would defect in any case since, irrespective of the strategy of the other,
the optimal strategy is to give $0$. Yet in day-to-day life people often do
cooperate and, perhaps unsurprisingly, research has shown that even in
anonymous one-shot PD experiments a significant percentage of people choose to
cooperate (see, e.g., Rapoport, 1965). 

\par More recently, behavioural scientists have delved into choices people make regarding deception in different circumstances and under different conditions. Unlike cooperation and altruism, lying aversion is not measured by a unique and standard economic game and (at least) three different models have been put forward (Gneezy, Rockenbach \& Serra-Garcia, 2013). However, irrespective of the model used, findings all point to the same direction: while the classic approach in economics assumes that people are selfish and that lying in itself does not involve any cost, accumulating evidence suggests that a significant amount of people are lie-averse in economic and social interactions (Abelar, Becker \& Falk, 2014; Cappelen, S\o rensen \& Tungodden 2013; Erat \& Gneezy, 2012;  Gneezy, 2005; Gneezy, Rockenbach \& Serra-Garcia, 2013; Hurkens \& Kartik, 2009; Lundquist, Ellingsen, Gribbe \& Johannesson, 2009; Weisel \& Shalvi, 2015). 

\par Recent research has shed light also on \emph{when}  people are more likely to use deception. Shalvi, Dana, Handgraaf and de Dreu (2011) find that ``observing desired counterfactuals attenuates the degree to which people perceive lies as unethical''. Wiltermuth (2011) finds that people are more likely to cheat when the benefit of doing so is split between themselves and another person, even when the other beneficiary is a stranger with whom they had no interaction.
Gino, Arial \& Ariely (2013) distinguish among the mechanisms that may drive this increased willingness to cheat when the spoils are split with others. They suggest that the ability to justify self-serving actions as appropriate when others benefit is a stronger driver for unethical behavior than pure concern for others. They also find that people cheat more when the number of beneficiaries increases and that individuals feel less guilty about their dishonest behavior when others benefit from it.
Conrads, Irlenbusch, Rilke \& Walkowitz (2013) examine the impact of two prevalent compensation schemes, individual piece-rates (under which each individual gets one compensation unit for each unit they produce) and team incentives (under which the production of the team is pooled and each individual receives one half of a compensation per unit of the joint production output). They find that lying is more prevalent under team incentives than under the individual piece-rates scheme. Thus, their results add to the evidence in Wiltermuth (2011) and Gino et al. (2013) suggesting that individuals are more willing to lie when the benefits of doing so are shared with others.
Cohen, Gunia, Kim-Jun and Murnighan (2009) test whether groups lie more than individuals. They find that groups are more inclined to lie than individuals when deception is guaranteed to best serve their economic interest, but lie relatively less than individuals when honesty can be used strategically. Their results suggest that groups are more strategic than individuals in that they will use or avoid deception in order to maximize their economic outcome.

\par
However, with few exceptions, no previous studies have investigated the relation between prosocial behaviour and aversion to telling white lies. Shalvi and de Dreu (2014) show that oxytocin, a neuropeptide known to promote affiliation and cooperation with others, promote group-serving dishonesty. Levine and Schweitzer (2014) report that people who tell weakly altruistic white lies (lies that benefit the other person at a small or even null cost for the liar) are perceived as more moral than those who tell the truth. In a subsequent work, Levine and Schweitzer (2015) show that trustors in a trust game allocate more money to people who have told a weakly altruistic white lie in a previous game than to people who have told the truth. This result provides evidence that telling a weakly altruistic white lies signal prosocial behavior in observers. However, Levine and Schweitzer (2015) do not measure trustees' behavior and thus it remains unclear whether those who tell a weakly altruistic white lies are really more prosocial than those who tell the truth. One corollary of the results of the current paper is a positive answer to this question. 

\par More closely related to our work is that of Cappelen et al. (2013), which explores the correlation between altruism in the DG and aversion to telling a Pareto white lie (PWL), that is, a lie that increases the benefits of both the liar and the listener, providing evidence that telling a Pareto white lie give significantly \emph{less} in the Dictator Game. Our work build on and extend this paper.

\par
Indeed, although its results represent a good starting point, more research is needed to develop a better understanding of the relation between aversion to telling a white lie and prosocial behaviour. First of all, most everyday situations are better modelled by a PD, rather than a DG. Since altruism in the DG and cooperation in the PD are different behaviours (virtually all altruistic people cooperate, but the converse does not hold - see Capraro, Jordan \& Rand, 2014), it is also important to investigate the correlation between cooperation in the PD and aversion to telling a white lie. Second, many white lies are not Pareto optimal, but involve a cost for the liar (altruistic white lies, AWL). Thus, it is important to go beyond Pareto white lies and explore also the correlation between prosocial behaviour and altruistic white lies. 

\par To fill this gap, we implemented a 2x2 experiment, in which subjects play a two-stage game. In the first stage they play one out of two possible treatments in a variation of the Deception Game introduced by Gneezy et al. (2013). In these treatments they have the opportunity to tell either a Pareto or an altruistic white lie. In the second stage participants are assigned to either the PD or the DG. We refer the reader to the next section for more details about the experimental design and to the Results section for a detailed description of the results. Here we anticipate that we have found evidence of three major results: (i) both altruism and cooperation are positively correlated with aversion to telling a Pareto white lie; (ii) both altruism and cooperation
are negatively correlated with aversion to telling an altruistic white lie; (iii) men are more likely than women to tell an altruistic white lie, but not to tell a Pareto white lie.

\section{Experimental design and procedure}

\par We set up a two-stage experiment in which we first collect data on participants' lying aversion; followed by data regarding their pro-social preferences. In the first stage, participants were directed to one of two variations on the Deception Game, in the spirit of Gneezy et al. (2013). One variation serves to measure aversion to tell an altruistic white lie; the other variation serves to measure aversion to tell a Pareto white lie. In the second stage of the experiment, the players were randomly assigned to either the DG or the PD. Comprehension questions were asked for each of the four games, before any decision could be made. Participants failing any of the comprehension questions were automatically excluded from the survey. In the next subsections we describe the experimental design. Full experimental instructions are reported in Appendix A.

\subsection{Stage 1: Measure of lying-aversion}
 
\par In the first stage of the experiment, participants played a Deception Game akin to that of Gneezy et al. (2013) with Pareto White Lies (PWL) and Altruistic White Lies (AWL) treatments. As in Gneezy et al. (2013) two players are paired and the first player has the opportunity to tell a white lie. However, unlike Gneezy et al. (2013), in our Deception game the payoffs of both players depend only on Player 1's choice and not on whether Player 2 believes that Player 1 is telling the truth or telling a lie. In our Deception Game, Player 2 is passive and does not make any decision. We use this variant because we are interested in looking at the relation between Player 1's lying aversion and their pro-social tendencies. Our design allows us to avoid confounding effects due to the beliefs that Player 1 may have about the beliefs of Player 2. Since in our design Player 2 does not make any decision, the beliefs that Player 1 may have about Player 2 do not make any role and a pro-social Player 1 will always tell a white lie, regardless of their beliefs. 

\par Specifically, in our Deception Game, Player 1 is assigned to group $i$, where $i\in\{1, 2\}$. The group allocation is communicated only to Player 1. Player 1 can choose between two possible strategies. Option A: telling the number of the group they were assigned to; or Option B: telling the number of the other group.  Players in the AWL condition were told that the payoff for each player would be determined as follows:
\begin{itemize}
\item Option A: Player 1 and Player 2 earn $\$0.10$ each. 
\item Option B: Player 1 earns $\$0.09$ and Player 2 earns $\$ 0.30$. 
\end{itemize}

\par Players in the PWL condition, on the other hand, were told that the payoff for each player would be determined as follows:
\begin{itemize}
\item Option A: Player 1 and Player 2 earn $\$ 0.10$ each. 
\item Option B: Player 1 and Player 2 earn $\$ 0.15$  each. 
\end{itemize}

\subsection{Stage 2: Measure of pro-sociality}

In the second stage of the game, all participants were randomly assigned to either a one-shot anonymous Dictator Game (DG) or a one-shot anonymous Prisoner's Dilemma (PD) to assess the extent of their altruism toward, or cooperation with, unrelated individuals.
\par In the DG, dictators were given an initial endowment of $\$ 0.20$ and were asked to decide how much money, if any, to \emph{transfer} to a recipient, who was given nothing. Each dictator was informed that the recipient they were matched with would have no active role and would only receive the amount of money the dictator decides to give. In the PD, participants were given an initial endowment of $\$ 0.10$ and were asked to decide whether to \emph{transfer} the $\$ 0.10$ to the other participant (cooperate) or not (defect). Each time a participant transfers their $\$0.10$, the other participant earns $\$0.20$. Each participant was informed that the participant they were matched with would be facing the same decision problem. 
\par We deliberately chose to use the word ''transfer'', rather than ``give'', ``cooperate'', or similar words, in order to minimise possible framing effects caused by the moral weight associated with names of the strategies.

\section{Results}
Participants living in the United States were recruited via the crowd-sourcing internet marketplace Amazon Mechanical Turk (Paolacci, Chandler \& Ipeirotis, 2010; Horton, Rand \& Zeckhauser, 2011; Bartneck, Duenser, Moltchanova \& Zawieska, 2015).  A total of 1212 subjects (59\% males, mean age = 33.83) passed the comprehension questions and participated in our experiment.  

\par In the first stage of the experiment, 614 subjects played in the AWL treatment while 598 subjects were assigned to the PWL treatment. Pareto white lies were told extremely more frequently than altruistic white lies: whilst only 23\% of the participants chose to tell an altruistic white lie, 83\% of the subjects lied in the PWL treatment (Wilcoxon Rank sum, $p < .0001$). These results are qualitatively in line with those reported by Erat and Gneezy (2012), who found that 43\% of people lie in their AWL treatment and 76\% of participants lie in their PWL treatment. The effect of demographic questions on lying aversion will be discussed separately.

\par In the second stage of the experiment, 697 participants were assigned to the DG, while 515 played the PD. Dictators on average transferred 22\% of their endowment, whilst in the PD cooperation was chosen 35\% of the time. Linear regression predicting DG donation as a function of the three main demographic variables (sex, age, and education level) shows that women donated more than men (coeff = $1.74$, $p < .0001$), that elders donated slightly more than younger people (coeff $=0.03$, $p=0.048$) and that education level has no significant effect on DG donations (coeff = 0.04, $p=0.78$)\footnote{The fact that women give more than men in the DG is reasonably well established, as the majority of studies report either this effect (Eckel \& Grossmann, 1997; Andreoni \& Vesterlund, 2001; Dufwenberg \& Muren, 2006; Houser \& Schunk, 2009; Dreber, Ellingsen, Johannesson \& Rand, 2013; Dreber, von Essen \& Ranehill 2014; Capraro \& Marcelletti 2014; Capraro 2015) or a null effect (e.g. Dreber et al. 2013; Bolton \& Katok, 1995). Also the fact that elders donate more than younger people is relatively well established (see Engel (2011) for a meta-analysis reporting a marginally significant effect and Capraro \& Marcelletti (2014) for a replication of this effect on an AMT sample).}. On the other hand, logit regression predicting cooperation as a function of the three main demographic variables shows that none of them has a significant effect on cooperative behaviour (all $p$'s $>0.15$). 

 \wl
\hspace{0.38 cm} \textbf{Altruism and lying-aversion} 
\par Figure 1 reports the average DG donation of liars and honests in both the AWL and the PWL treatments and suggests that honest people were more altruistic than liars in the PWL treatment, but less altruistic than liars in the AWL treatment. To confirm this we use linear regression predicting DG donation using a dummy variable, which takes value 1 (resp. 0) if the participant has told the truth (resp. a lie). Results show that, in the AWL treatment, honest people were marginally significantly less altruistic than liars (coeff $=-1.14$, $p = 0.063$), and that, in the PWL treatment, honest people were significantly more altruistic than liars (coeff $= 1.43$, $p = 0.035$). Next we examine whether these differences are driven by individual differences. To do this, we repeat the linear regressions including controls on the three main demographic variables (sex, age, and level of education). Results show that, in the AWL treatment, honest people were significantly less altruistic than liars (coeff $=-1.33$, $p = 0.03$), and that, in the PWL treatment, honest people were marginally significantly more altruistic than liars (coeff $ = 1.26$, $p = 0.06$). In both cases, the only significant demographic variable is the gender of the participant (AWL: coeff $= 1.74$, $p = 0.0008$, PWL: coeff $= 1.79$, $p = 0.0008$). Full regression table is reported in Appendix B, Table 1. Thus, although the difference in altruism between liars and honest people is partly driven by the gender of the participant, it remains significant or close to significant also after controlling for this variable, suggesting existence of a true effect of aversion to tell a white lie on altruistic behaviour in the Dictator Game. 
\begin{figure}[h] 
   \centering
   \includegraphics[scale=0.70]{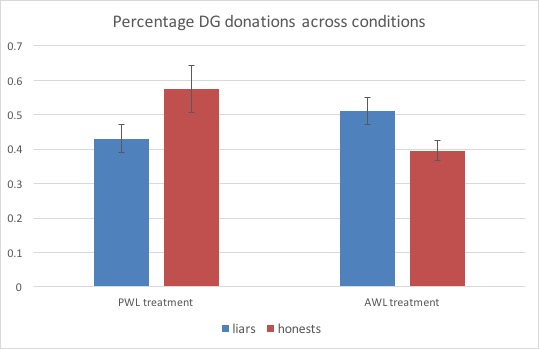} 
   \caption{\emph{Average DG donation of liars and honests in both the AWL and the PWL treatments. Error bars represent the standard errors of the means. In the Pareto White Lies treatment, honest people tend to me more altruistic than liars (linear regression with no control on socio-demographic variables: coeff $= 1.43$, $p = 0.035$; with control: coeff $= 1.26$, $p = 0.06$). In the Altruistic White Lies treatment, honest people tend to me less altruistic than liars (linear regression with no control on socio-demographic variables: coeff $= -1.14$, $p = 0.063$; with control: coeff $= -1.33$, $p = 0.03$).}}
   \label{fig:DGlie}
\end{figure}

\wl
\hspace{0.38 cm}\textbf{Cooperation and lying-aversion} 
\par \par Figure 2 reports the average PD cooperation of liars and honests in both the AWL and the PWL treatments and suggests that, as in the DG case, honest people were more cooperative than liars in the PWL treatment, but less cooperative than liars in the AWL treatment. To confirm this we use logit regression predicting PD cooperation using a dummy variable, which takes value 1 (resp. 0) if the participant has told the truth (resp. a lie). Results show that, in the AWL treatment, honest people were highly less altruistic than liars (coeff $=-1.25$, $p < .0001$), and that, in the PWL treatment, honest people were significantly more altruistic than liars (coeff $= 0.71$, $p = 0.04$). Next we examine whether these differences are driven by individual differences. To do this, we repeat the logit regressions including controls on the three main demographic variables. Results show that, in the AWL treatment, honest people were less altruistic than liars (coeff $=-1.31$, $p < .0001$), and that, in the PWL treatment, honest people were more altruistic than liars (coeff $ = 0.79$, $p = 0.02$). In both cases, none of the demographic variable is significant (only gender has a marginally significant effect ($p = 0.08$), but only in the PWL treatment). Full regression table is reported in Appendix B, Table 2. 

\begin{figure}[h] 
   \centering
   \includegraphics[scale=0.70]{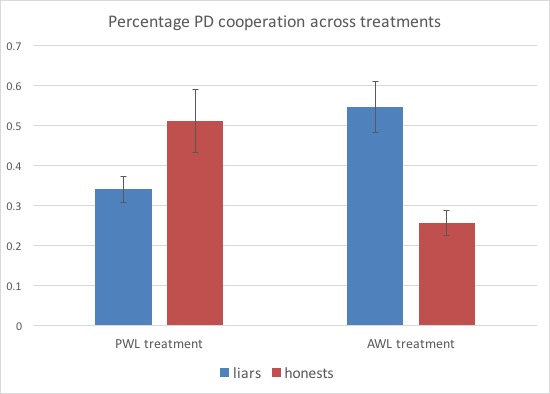} 
   \caption{\emph{Average PD cooperation of liars and honests in both the AWL and the PWL treatments. Error bars represent the standard errors of the means. In the Pareto White Lies treatment, honest people tend to me more cooperative than liars (logit regression with no control on socio-demographic variables: $= 0.71$, $p = 0.04$; with control: coeff $ = 0.79$, $p = 0.02$). In the Altruistic White Lies treatment, honest people tend to me less cooperative than liars (logit regression with no control on socio-demographic variables: coeff $=-1.25$, $p < .0001$; with control: coeff $=-1.31$, $p < .0001$).}}
   \label{fig:PDlie}
\end{figure}

\wl
\hspace{0.38 cm}\textbf{Gender differences in deception}

\par Gender differences in deceptive behaviour have attracted considerable attention since the work of Dreber and Johannesson (2008), who found that men are more likely than women to tell a black lie, that is, a lie that benefits the liar at the expenses of the listener. In the context of white lies, Erat and Gneezy (2012) found that women are more likely than men to tell an altruistic lie, but men are more likely than women to tell a Pareto white lie. Interestingly, the latter result was not replicated by Cappelen et al. (2013), who found no gender differences in lying aversion in the context of Pareto white lies. In line with this latter result, we also find no gender differences in the PWL treatment. Indeed, logit regression predicting the probability of telling a Pareto white lie as a function of sex, age, and level of education shows no significant effect of gender (coeff $= 0.25$, $p = 0.25$) and age (coeff $=-0.01$, $p = 0.47$) and, if anything, shows a significant negative effect of the level of education (coeff $=-0.18$, $p = 0.04$). Interestingly, in the context of altruistic white lies, we even find the reverse correlation  of that reported in Erat and Gneezy (2012). In our sample, men are slightly more likely than women to tell an altruistic white lie (26\% vs 18\%). The difference is statistically significant as shown by logit regression predicting the probability of telling an altruistic white lie as a function of sex, age, and level of education (gender: coeff $= 0.50$, $p=0.02$; age: coeff $=-0.00$, $p = 0.85$; education: coeff $=-0.02$, $p = 0.73$). We refer the reader to Figure 3 for a visual representation of gender differences in deceptive behaviour.

\begin{figure}[h] 
   \centering
   \includegraphics[scale=0.70]{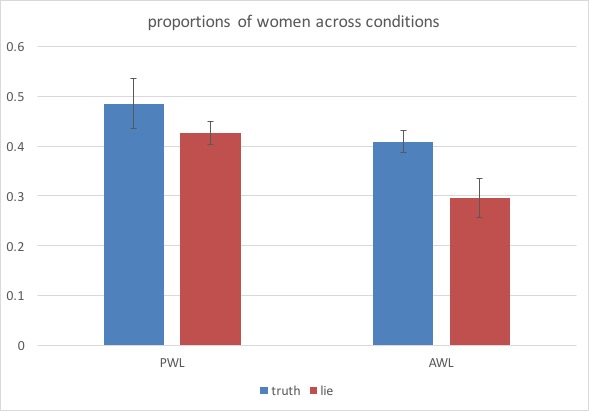} 
   \caption{\emph{Proportion of females across treatments divided between liars and honests. In the Pareto White lie treatment, there is no statistically significant gender difference in deceptive behaviour. On the other hand, in the Altruistic White lie treatment, we found that women are significantly more likely than men to tell the truth.}}
   \label{fig:PDlie}
\end{figure}

\section{Discussion}
\par We conducted this experiment to explore the relation between cooperation, altruism, and aversion to telling white lies among participants. Cooperative tendencies were measured through the Prisoner's Dilemma (PD); altruistic tendencies were measured through the Dictator Game (DG); and lying-aversion was measure using a Deception Game. The setup of our Deception Game was such that if Player 1 chose to lie then there would be an increase in monetary outcome for both players (the Pareto white lie variant, PWL) or an increase in monetary outcome for Player 2 at a small cost to Player 1 (the altruistic white lie variant, AWL). Our design differed from previous versions of the Deception Game in that the payoffs of both players depend only on the decision of the first player. This design allows us to study the correlation between Player 1's lying-aversion and pro-social tendencies without adding the potentially confounding factor that Player 1 may have beliefs about the behaviour of Player 2. 

\par Our results provide evidence of three major findings:  (i) both altruism and cooperation are positively correlated with aversion to telling a Pareto white lie; (ii) both altruism and cooperation are negatively correlated with aversion to telling an altruistic white lie; (iii) men are more likely than women to tell an altruistic white lie, but not to tell a Pareto white lie.

\par These results make several contributions to the literature. The positive correlation between altruism and aversion to telling a Pareto white lie was also found by Cappelen et al. (2013). Our results replicate and extend this finding as they also show that the same correlation holds true when considering cooperative behaviour (as opposed to altruistic behaviour), and that these correlations disappear and are actually even reversed when considering altruistic white lies (as opposed to Pareto white lies). These are not trivial extensions. Indeed, a positive correlation between altruism in the DG and aversion to telling a Pareto white lie can, in principle, be explained by assuming that there are two types of agents: (i) \emph{non-purely} utilitarian agents, who aim at maximising the social welfare and choose the strategy that maximises their payoff in case multiple strategies give rise to the same social welfare (e.g., Charness \& Rabin, 2002; Capraro, 2013); and (ii) \emph{purely} egalitarian agents, who minimise payoff differences, irrespective of their own payoff (e.g., Fehr \& Schmidt, 1999; Bolton \& Ockenfels, 2000 - with suitable values for the parameters of the models). Assuming this types distribution, non-purely utilitarian agents always tell Pareto white lies (because they increase the social welfare) and give nothing in the DG (because giving does not increase the social welfare); and purely egalitarian agents give half in the DG, yet are indifferent between telling a Pareto white lie or not in the Deception Game (since they both minimise payoff differences). Thus, if the proportion of utilitarian agents is large enough, this would generate a positive correlation between altruism in the DG and aversion to tell a Pareto white lie, that would be consistent with the findings in Cappelen et al. (2013). 

\par On the other hand, explaining our results using distributional preferences is much harder. Indeed, to explain the negative correlation between cooperation in the PD and telling a PWL, one must assume that the majority of utilitarian people actually defects in the PD. But this assumption clashes with the very nature of utilitarian people that is that of maximizing the total welfare and thus to cooperate in the PD.

\par One potential explanation for our findings is that subjects have two possible degrees of moral motivations (low or high) towards either of two moral principles (utilitarianism and deontology). Utilitarian people follow distributional preferences for maximizing the social welfare; deontological people follow non-distributional preferences for doing what they think it is the right thing to do. We assume that these types of individuals act as follows:

\commentout{
explanation is that a proportion of people
 experience a conflict between maximising their own monetary payoff and
 doing what they think it is the morally right thing to do, whose proximate
 path is a desire to avoid norm-based guilt (Basile and Mancini, 2011; Carn\`i, Petrocchi, Del Miglio, Mancini and Couyoumdjian, 2013; Izard, 1977; Lewis,
 1971; Monteith, 1993; Mosher, 1965; Piers and Singer, 1971; Wertheim and
 Schwartz, 1983). According to this interpretation, a proportion of these
 "homines morales" may have non-distributional notions of what they think it
 is the morally right thing to do: irrespective of economic consequences,
 they cooperate, share, tell the truth. Besides explaining our first finding, interestingly, this interpretation is consistent also with our second finding. On the one hand, telling the truth in the AWL condition maximises the sender's individual gain and so attracts self-regarding agents; on the other hand, telling an altruistic white lie maximises the gain of the receiver at a cost to the liar and so attracts other-regarding agents. These attractors work in opposite directions to those observed in the PWL conditions and might explain the reversal of significant interaction between pro-sociality and lying aversion in the AWL treatments.
}

\begin{itemize}
\item \emph{High utilitarian} people give half in the DG, cooperate in the PD, and lie in the AWL and in the PWL.
\item \emph{High deontological} people give half in the DG, cooperate in the PD, and tell the truth in the AWL and in the PWL.
\item \emph{Low utilitarian} people keep in the DG, keep in the PD, tell the truth in the AWL and are indifferent in the PWL.
\item \emph{Low deontological} people keep in the DG, keep in the PD, tell the truth in the AWL and are indifferent in the PWL.
\end{itemize}

According to this partition, the positive correlation between truth-telling and DG-donation/PD-cooperation in the PWL treatment would be driven by \emph{high deontological} subjects; and the negative correlation between truth-telling and DG-donation/PD-cooperation in the AWL treatment would be driven by \emph{low utilitarian} and \emph{low deontological} subjects.

\par There might be \emph{high utilitarian }subjects as well, but they do not show
up because we have only one case per subject. An interesting direction for future research is therefore to do a within-subject design with many trials, using different payoffs, aimed at establishing the position of each subject in a two-dimensional space. 

\par Of course, more research is needed also to support the existence of a possibly non-distributional ``deontological domain'', containing all those actions that a
 particular individual considers to be morally right independently of their economic consequences, and to classify the actions belonging to this domain. For instance, here we have focussed on altruism and cooperation, as they are the most studied pro-social behaviours. However, they are not the only ones. Future research may be devoted to understand whether the same correlations with truth-telling hold for other pro-social behaviours, such as benevolence (i.e., acting in such a way to increase the other's payoff beyond one's own, Capraro, Smyth, Mylona, \& Niblo 2014) and hyper-altruism (i.e., weighting the other's payoff more than one's own, Crockett, Kurth-Nelson, Siegel, Dayan \& Dolan 2014; Capraro 2015). 

\par Additionally, our results connect to the work of Levine and Schweitzer (2015), which reported that telling an altruistic white lie signals pro-social tendencies in observers: third parties, when playing in the role of trustors in the Trust Game, allocate more money to people who have told an altruistic white lie in a previous Deception Game than to those who have told the truth. However, Levine and Schweitzer (2015) did not measure trustees' behavior and so it remained unclear whether people telling an altruistic white lie were really more prosocial than those telling the truth, or it was an observer's false belief. Our results provide evidence that this is not a false belief, as they show that subjects telling a white lie are both more altruist and more cooperative than those telling the truth.

\par Finally, our results add to the literature regarding gender differences in deceptive behaviour. Dreber and Johannesson (2008) found that men were more likely than women to tell black lies (e.g., lies that increase the liar's benefit at the expense of the listener). A similar result was shown by Friesen and Gangadharan (2012), who found that men are more likely than women to behave dishonestly for their own benefit. Yet, Childs (2012) failed to replicate this gender effect using a very similar design to that in Dreber and Johannesson (2008). In the context of white lies, Erat and Gneezy (2012) reported that women are more likely than men to tell an altruistic white lie, but men are more likely than women to tell a Pareto white lie. This latter result was not replicated by Cappelen et al. (2013), who found no gender differences in telling a Pareto white lie. In line with the latter result, our results also show no gender difference in telling a Pareto white lie. But, interestingly, we found gender differences in telling an altruistic white lie, but in the opposite direction than that reported in Erat and Gneezy (2012): we found that men are more likely than women to tell an altruistic white lie. Taken together, these results suggest that it may be premature to draw general conclusions about whether there are general gender differences in lying, and call for further studies.

\pagebreak
\begin{appendices}

\section{Experimental instructions}

\subsection{Deception Game: Altruistic White Lies treatment}

\par\medskip Please read these instructions carefully. You may earn a considerable sum of money, depending on the decisions you make in the experiment.
 
\par\medskip You have been matched with another participant.  Your bonus and that of the other participant depends only on your choice. The other participant does not play any active role.
 
\par\medskip There are two groups (group 1, group 2). You will be randomly assigned to one of them. You will be informed of the group you have been assigned to, but the other participant will not. 
 
\par\medskip We will ask you to declare the group you have been assigned to. So you get to choose between two possible options:

\begin{itemize} 
\item Option 1: I have been assigned to group 1.
\item Option 2: I have been assigned to group 2. 
\end{itemize} 

\par\medskip If you choose the option corresponding to the number of the actual group you have been assigned to, then you will receive 10c and the other participant will receive 10c.

\par\medskip If you choose the option correponding to the number of the other group (the one you have not been assigned to), then you will receive 9c and the other participant will receive 30c.
 
\par\medskip Finally, only you will be informed of the particular monetary value connected to each message. The other participant will not be informed of these monetary values.

\par\medskip Here are some questions to ascertain that you understand the rules. Remember that you have to answer all of these questions correctly in order to get the completion code. If you fail any of them, the survey will automatically end and you will not get any payment.

\par\medskip What is the choice that maximise YOUR outcome? \emph{Available answers}: Choosing the message corresponding to the number of the actual group you have been assigned to/Choosing the message corresponding to the number of the other group (the one you have not been assigned to).

\par\medskip What is the choice that maximise the OTHER PARTICIPANT'S outcome? \emph{Available aswers}: Choosing the message corresponding to the number of the actual group you have been assigned to/Choosing the message corresponding to the number of the other group (the one you have not been assigned to).

\par\medskip Congratulations, you have passed all comprehension questions. It is time to make your real choice.

\emph{Here participants were randomly divided in two conditions, corresponding to the two possible groups. We report the instructions only for Group 1.}

\par\medskip You have been assigned to group 1. 

\par\medskip Which option do you choose?

\subsection{Deception Game: Pareto White Lies treatment}

\par\medskip Please read these instructions carefully. You may earn a considerable sum of money, depending on the decisions you make in the experiment.
 
\par\medskip You have been matched with another participant.  Your bonus and that of the other participant depends only on your choice. The other participant does not play any active role.
 
\par\medskip There are two groups (group 1, group 2). You will be randomly assigned to one of them. You will be informed of the group you have been assigned to, but the other participant will not. 
 
\par\medskip We will ask you to declare the group you have been assigned to. So you get to choose between two possible options:

\begin{itemize} 
\item Option 1: I have been assigned to group 1.
\item Option 2: I have been assigned to group 2. 
\end{itemize} 

\par\medskip If you choose the option corresponding to the number of the actual group you have been assigned to, then you will receive 10c and the other participant will receive 10c.

\par\medskip If you choose the option correponding to the number of the other group (the one you have not been assigned to), then you will receive 15c and the other participant will receive 15c.
 
\par\medskip Finally, only you will be informed of the particular monetary value connected to each message. The other participant will not be informed of these monetary values.

\par\medskip Here are some questions to ascertain that you understand the rules. Remember that you have to answer all of these questions correctly in order to get the completion code. If you fail any of them, the survey will automatically end and you will not get any payment.

\par\medskip What is the choice that maximise YOUR outcome? \emph{Available aswers}: Choosing the message corresponding to the number of the actual group you have been assigned to/Choosing the message corresponding to the number of the other group (the one you have not been assigned to).

\par\medskip What is the choice that maximise the OTHER PARTICIPANT'S outcome? \emph{Available aswers}: Choosing the message corresponding to the number of the actual group you have been assigned to/Choosing the message corresponding to the number of the other group (the one you have not been assigned to).

\par\medskip Congratulations, you have passed all comprehension questions. It is time to make your real choice.

\emph{Here participants were randomly divided in two conditions, corresponding to the two possible groups. We report the instructions only for Group 1.}

\par\medskip You have been assigned to group 1. 

\par\medskip Which option do you choose?

\subsection{Dictator Game}

Please read these instructions carefully. You may earn a considerable sum of money, depending on the decisions you make in the experiment.

\par\medskip You have been paired with another participant. The amount of money you can earn depends only on your choice.

\par\medskip You are given $20c$ and the other participant is given nothing.

\par\medskip You have to decide how much, if any, to transfer to the other participant.

\par\medskip The other participant has no choice, is REAL, and will really accept the amount of money you decide to transfer.

\par\medskip No deception is used. You will really get the amount of money you decide to keep.

\par\medskip Here are some questions to ascertain that you understand the rules. Remember that you have to answer all of these questions correctly in order to get the completion code. If you fail any of them, the survey will automatically end and you will not get any payment. 

\par\medskip What is the transfer by you that maximizes your bonus? \emph{Available aswers}: 0c/2c/4c/.../20c.

\par\medskip What is the transfer by you that maximizes the other participant's bonus? \emph{Available aswers}: 0c/2c/4c/.../20c.

\par\medskip Congratulations, you have answered both comprehension questions correctly! 

\par\medskip It is now time to make your choice.

\par\medskip What amount will you transfer to the other person? \emph{Available options}: 0c/2c/4c/.../20c.

\subsection{Prisoner's Dilemma}

Please read these instructions carefully. You may earn a considerable sum of money, depending on the decisions you make in the experiment.

\par\medskip You have been paired with another anonymous participant. You are both given $10c$ and each of you must decide whether to transfer the $10c$ or not. Each time a participant transfers their $10c$, the other participant earns $20c$.

\par\medskip So: \begin{itemize}

\item If you both decide to transfer the 10c, you end the game with 20c

\item If the other participant transfers the 10c and you do not, you end the game with 30c

\item If you transfer the 10c and the other participant does not, you end the game with 0c

\item If neither of you transfer the 10c, then you end the game with 10c

\end{itemize}

\par\medskip Here are some questions to ascertain that you understand the rules. Remember that you have to answer all of these questions correctly in order to get the completion code. If you fail any of them, the survey will automatically end and you will not get any payment. 

\par\medskip What choice should you make to maximise your gain? \emph{Available aswers}: Transfer the 10c/Do not tranfer the 10c.

\par\medskip What choice should you make to maximise the other participant's gain? \emph{Available aswers}: Transfer the 10c/Do not tranfer the 10c.

\par\medskip What choice should the other participant make to maximise your gain? \emph{Available aswers}: Transfer the 10c/Do not tranfer the 10c.

\par\medskip What choice should the other participant make to maximise their gain? \emph{Available aswers}: Transfer the 10c/Do not tranfer the 10c.

\par\medskip Congratulations, you have answered both comprehension questions correctly! 

\par\medskip It is now time to make your choice. \emph{Available options}: Tranfer the 10c/Do not trasfer the 10c

\pagebreak
\section{Regression tables}

\begin{center}

\begin{tabular}{| l | c | c | c | c | c | c | c |}
   \hline         
  &I&II&III&IV&V&VI\\
  \hline         
 AWL &-1.14*&&-1.33*&&&\\
&(0.61)&&(0.60)&&&\\
\hline
 PWL &&&&1.43*&&1.26*\\
&&&&(0.68)&&(0.67)\\
\hline
  sex &  &1.63**& 1.74*** && 1.87*** & 1.80*** \\
   & &(0.51)& (0.51) && (0.53) & (0.53) \\
\hline
  age & &0.03& -0.03&& -0.03 & 0.03 \\
 &  &(0.02)& (0.02) && (0.02) & (0.02) \\
\hline
  education &  &0.01&  0.01 && 0.07 & 0.11  \\
 &  &(0.21)& (0.21) && (0.21)  & (0.21) \\
\hline
  constant & 5.10*** &0.61& 1.54&4.31***& 0.58 &0.31\\
  & (0.54)&(1.38)& (0.28) &(0.29)& (1.39) &(1.39)\\
\hline
  No. cases & 357 &357& 357&340 &340& 340 \\
  \hline  
\end{tabular}\captionof{table}{Summary of the statistical analysis regarding the Dictator Game. We ran linear regression predicting DG donation. The explanatory variable \emph{AWL} (resp. \emph{PWL}) takes value 0 if a participant lied in the AWL (resp. PWL) treatment, and value 1 otherwise. The explanatory variable \emph{sex} takes value 1 (resp. 2) if a participant is a man (resp. woman). We report coefficient, standard error (in brackets, below the coefficient), and significance levels using the notation: *: $p<0.1$, **: $p<0.01$, and ***: $p<0.001$.}
 \end{center}

\pagebreak

\begin{center}

\begin{tabular}{| l | c | c | c | c | c | c | c |}
   \hline         
  &I&II&III&IV&V&VI\\
  \hline         
 AWL &-1.25***&&-1.31***&&&\\
&(0.30)&&(0.31)&&&\\
\hline
 PWL &&&&0.70*&&0.78*\\
&&&&(0.34)&&(0.35)\\
\hline
  sex &  &0.03& 0.14 && 0.45* & 0.46* \\
   & &(0.28)& (0.29) && (0.26) & (0.27) \\
\hline
  age & &0.01& -0.01&& -0.01 & 0.01 \\
 &  &(0.01)& (0.01) && (0.01) & (0.01) \\
\hline
  education &  &-0.14&  -0.16 && 0.08 & 0.10  \\
 &  &(0.12)& (0.12) && (0.11)  & (0.12) \\
\hline
  constant & 0.19 &-0.36& 0.48&-0.65& -1.83* &-2.17\\
  & (0.25)&(0.63)& (0.81) &(0.14)& (0.73) &(0.76)\\
\hline
  No. cases & 257 &257& 257&258 &258& 258 \\
  \hline  
\end{tabular}\captionof{table}{Summary of the statistical analysis regarding the Prisoner's Dilemma. We ran logit regression predicting PD cooperation (0 = defection, 1 = cooperation). The explanatory variable \emph{AWL} (resp. \emph{PWL}) takes value 0 if a participant lied in the AWL (resp. PWL) treatment, and value 1 otherwise. The explanatory variable \emph{sex} takes value 1 (resp. 2) if a participant is a man (resp. woman). We report coefficient, standard error (in brackets, below the coefficient), and significance levels using the notation: *: $p<0.1$, **: $p<0.01$, and ***: $p<0.001$.}
 \end{center}

\end{appendices}

\end{document}